# Electroencephalography source connectivity: toward high time/space resolution brain networks

Hassan M.[1,2] and Wendling F.[1,2]

[1] University of Rennes 1, LTSI, F-35000 Rennes, France
[2] INSERM, U1099, F-35000 Rennes, France

*Abstract*— The human brain is a large-scale network which function depends on dynamic interactions between spatially-distributed regions. In the rapidly-evolving field of network neuroscience, two yet unresolved challenges are potential breakthroughs. First, functional brain networks should be estimated from noninvasive and easy to use neuroimaging techniques. Second, the time/space resolution of these techniques should be high enough to assess the dynamics of identified networks. Emerging evidence suggests that Electroencephalography (EEG) source connectivity method may offer solutions to both issues provided that scalp EEG signals are appropriately processed. Therefore, the performance of EEG source connectivity method strongly depends on signal processing (SP) that involves various methods such as preprocessing techniques, inverse solutions, statistical couplings between signals and network science. The main objective of this tutorial-like review is to provide an overview on EEG source connectivity. We describe the major contributions that the SP community brought to this research field. We emphasize the methodological issues that need to be carefully addressed to obtain relevant results and we stress the current limitations that need further investigation. We also report results obtained in concrete applications, in both normal and pathological brain states. Future directions in term of signal processing methods and applications are eventually provided.

# I. INTRODUCTION

Over the past decades, neuroscience research has significantly improved our understanding of the normal brain. There is now a growing body of evidence suggesting that brain functions are generated by large-scale networks of highly specialized and spatially segregated areas of the nervous system. From a theoretical viewpoint, network science in general and graph theory in particular has progressively entered the fields of neuroscience and neurology. A relatively new research field, referred to as "*network neuroscience*" [1] offered researchers a unique opportunity to assess, quantify and ultimately understand the multi-faceted features of complex brain networks. This inter-disciplinarily field has also been accelerated by the enormous advances of neuroimaging techniques that now allow for visualization of brain structure and function at unprecedented space and time resolution using, for instance, functional Magnetic Resonance Imaging -fMRI-, Magneto-Encephalography -MEG- or Electro-Encephalography -EEG- .

In this rapidly-growing and thought-provoking context, the identification of normal and pathological functional networks from neuroimaging data has become one of the most promising prospects in brain research [2]. Among the neuroimaging techniques able to provide relevant information about the dynamics of functional brain networks, EEG has considerably progressed over the two past decades. A key advantage of EEG systems is the non-invasiveness and the relative easiness of use. Information conveyed in EEG signals can be highly informative about the underlying functional brain networks, if those signals are appropriately processed to extract relevant information. In addition, an important advantage of EEG is the excellent temporal resolution that offers the irreplaceable opportunity not only to track large-scale brain networks over very short duration which is the case in many cognitive tasks [3], but also to analyse fast dynamical changes that can occur during resting state [4] or in brain disorders like epilepsy, typically during interictal (period between seizures) or ictal (during seizure) events.

The role of neural synchrony in brain functions, using EEG, has been reviewed in depth [5]. Most of the reported studies on functional connectivity analyses from EEG were performed at the sensor level. However, the interpretation of corresponding networks is not straightforward as signals are strongly corrupted by the volume conduction effect due to the electrical conduction properties of the head [6, 7] and the fact that multiple scalp electrodes collect, to some extent, the activity arising from the same brain sources. These two factors can result in an inaccurate estimation of the real functional connectivity between brain areas. Several recent studies have clearly reported the limitations of computing connectivity at the EEG scalp level, see [8] for review. The recent past years have witnessed a significant increase of interest for EEG analysis of functional brain networks at the level of cortical sources. The proposed approach is called *EEG source connectivity*. While reducing the above limitations, it is also conceptually very attractive as high spatiotemporal resolution networks can be directly identified in the cortical source space provided that some methodological aspects are carefully accounted for to avoid pitfalls.

Practically, the transition from the electrode space into source space involves solving an ill-posed inverse problem, the biophysical basis of which relying on the dipole theory. Among the many inverse methods proposed so far (review in [9]), some make use of physiologically-relevant a priori knowledge about both the location and the orientation of dipole sources at the origin of signals collected at the scalp. When this information is combined with accurate, possibly subject- or patient-specific, representation of the volume conductor (realistic head model [10] obtained by MRI segmentation), these methods considerably increase both the precision of localized sources and the estimation of associated time-series which are analogous to local field potentials. These time-series then become the input information to so-called "connectivity methods" which aim at estimating brain networks directly in the source space. It is worth noting that such networks are much more informative from the application viewpoint (cognitive sciences, clinics) [11, 12].

*EEG source connectivity* approaches involve several steps, each related to important topics in signal processing (SP) such as pre-processing of raw EEG data (artifact removal, denoising), EEG inverse solutions (source localization and reconstruction, spatial/temporal hypothesis, sparsity, regularization constraints…), estimation of statistical couplings between signals (phase synchronization/entropy, mutual information, coherence function, linear/nonlinear regression analysis) and graph-theory-based analysis (network segregation/integration and hubness). However, a complete overview of EEG source connectivity in term of methodological choices/limitations at each stage and the available tools is still missing. The main objective of this review is to address this issue by providing a comprehensive description of the main contributions from the SP community to this relatively new research field. From the methodological viewpoint, future advances which could likely overcome some limitations of current techniques are also addressed.

From the application viewpoint, we focus on results obtained so far with EEG source connectivity methods applied to data recorded during either normal or pathological brain states. We also present new results using this method in the tracking of the dynamics of brain networks during cognitive activity at sub-second time scale. In particular, we highlight the recent studies reporting attempts to use EEG source connectivity in order to reveal clinically valuable information about the topology and dynamics of dysfunctional networks involved in epilepsies and neurodegenerative diseases. Finally, some expectations in the field of cognitive and clinical research are also addressed.

The structure of this review is as follows: first we introduce the problem statement. We then describe the different steps performed from signal recordings to cortical brain networks with the associated methodological considerations and the available tools. Next we present the different applications of this technique to reveal brain networks involved in normal and pathological brain functions. We finally end with a discussion about the different limitations and the possible future directions.

## II. THE VOLUME CONDUCTION PROBLEM

Let **X(t)** be the time series recorded at the surface of the brain using **M** scalp EEG electrodes. These **M** sensors record the activity of **N** brain sources **S**(t). The computation of the statistical couplings directly between the **X(t)** time series produces **M×M** dimensional functional network at the scalp level. The scalp-EEG-based networks were widely used in the past [5]. However, interpretation of connectivity from sensor level recordings is very difficult, as these recordings are severely corrupted by the effects of 'field spread' and 'volume conduction' [6-8]. Ideally, if each electrode only measures the neuronal activity below the electrode then any statistical coupling measured from signals recorded from two electrodes $X_1$ and $X_2$ would reflect the connectivity between two physically-distinct brain regions $S_1$ and $S_2$ (Figure 1A). Unfortunately, this ideal situation cannot be always assumed for EEG recordings. Indeed, it is well known from the biophysics of the forward problem of EEG that each scalp electrode measures the activity arising from all brain sources, at a certain degree, depending on i) the source-to-sensor distance and ii) the orientation of equivalent dipoles associated with these sources. Therefore, scalp EEG signals correspond to a complex mixture of overlapping signals arising distinct brain regions. A direct consequence is statistical couplings measured in the electrode space (whatever the SP method used to this aim) cannot be interpreted, in a straightforward manner, as a brain connectivity measure between the underlying cortical regions (figure 1B).

Several methods have been proposed to deal with the volume conduction problem when computing connectivity at scalp level, such as the use of a spatial filter prior to computing connectivity (Laplacian montages), the computation of the time-lagged connectivity that would reflect a propagation process between distant areas or the use of measures less sensitive to volume conduction such as the imaginary part of the coherence. However, none of the proposed methods has shown to completely overcome the limitations of the volume conduction and the field spread problems, see [8] for more details about all the above mentioned approaches.

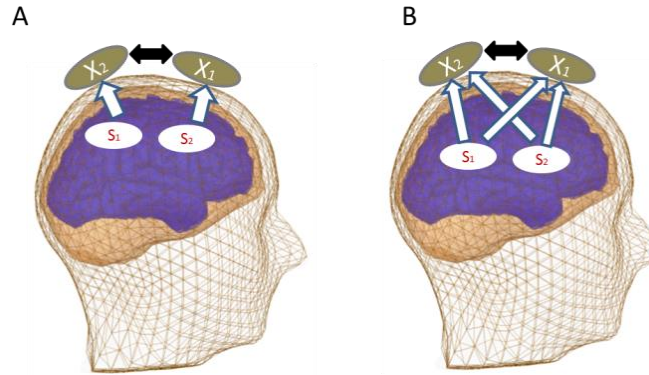

**Fig. 1.** *Illustration of the volume conduction problem for interpreting scalp-level connectivity.* **$X_1$ and $X_2$ represents the electrodes, $S_1$ and $S_2$ represent the brain sources, the black arrow represents the measured functional connectivity between $X_1$ and $X_2$ and the white arrow represents the pathway of electrical activity from $S_1$ and $S_2$. A) Ideally, each electrode measures brain activity below the electrode and thus connectivity between electrodes reflects connectivity between distinct brain regions, B) In practice, both brain sources $S_1$ and $S_2$ contribute to signals recorded at each electrode. Due to this mixing phenomenon, the statistical couplings measured in the electrode space cannot be directly interpreted in term of brain connectivity between the underlying cortical regions.**

### III. FROM EEG SIGNALS TO CORTICAL NETWORK

The computation of statistical couplings between EEG cortical sources reconstructed from the M channels is one of the most adequate method, so far, to reduce the volume conduction problem as the connectivity is computed at the level of the sources S(t). This can produce a network (at the cortical level) of N×N sources. Practically, this network is often reduced to R×R brain regions where R represents the number of regions of interests (ROI), which can vary depending on the segmentation parameters for the cortical surface (this issue will be considered in the next sections). This method, called "EEG source connectivity", is the main topic of this tutorial-like review.

The key idea of the EEG source connectivity method is the reconstruction of functional networks at the neocortical level from scalp recordings (see figure 2). The full pipeline from EEG recordings to cognitive/clinical markers of brain (dys)functions involves four main steps which are detailed in the following sections.

*A. Data recording and pre-processing*

  EEG data can be recorded during task-related or task-free paradigm (resting state). Depending on the context (clinical or cognitive research), these recordings can be performed using dense electrode arrays (64 to 256 sensors) either in patients or in healthy subjects (figure 1A). It is worth mentioning that MEG and EEG are very close techniques. From a biophysics viewpoint, the phenomena that are at the origin of recorded electric and magnetic fields are slightly different (EEG detects both radial and tangential currents, while MEG detects tangential currents only). Beside cost issues that directly stem from the technical difficulty of measuring magnetic fields in the order of 1 fT (10-12 Tesla), differences are also related to the sensitivity of both methods to deep sources, to the impact of volume conductor modelling on the reconstruction of sources and to the easiness of use. In this review we will be focusing on the EEG source connectivity method. Nevertheless, the analysis steps remain the same for both techniques.

*1) Number of channels*

The number of scalp electrodes is a crucial parameter for the performance of EEG source connectivity methods. Different studies showed that the number of channels has a high impact on the quality of the localized sources [13] or the networks reconstructed from scalp EEG data [14]. The use of the available systems (going from the former 19-32 to the newer 64-256 channels) can dramatically impact the performance of the source reconstruction step (see section B below). There is growing evidence that increasing the number of EEG channels provides greater accuracy in source estimation. The minimal number of electrodes required is also related to the other parameters used in the pipeline mainly the algorithm used to reconstruct the dynamics of brain sources. Many studies showed that at least 128 electrodes are needed to get satisfactory results, typically when the minimum norm class of inverse methods is used for localizing sources [13] or identifying functional networks [14].

*2) Pre-processing*

EEGs are often contaminated by various physiological or non-physiological sources of activity like, for instance, cardiac signals, eye movements/blinks, muscle activity or head/cable movements, among others. Removing these artifacts is a crucial step to produce "noise-free" signals prior to applying EEG source connectivity per se. The detection can be done visually, semi or fully automatically depending on the type of artifact. A simple way is to reject the segment of data where the artifact is visually clear. This is for instance the case for movement artifacts (participant moving head during an experiment) that simultaneously affect a high number of channels over a given time period. This step is still subjective as the visual inspection is user-dependent.

Artifacts can also be detected and removed automatically. The simplest method consists in comparing the EEG signal amplitude to an arbitrarily-defined threshold signal in order to remove non-physiological often saturated segments of very high amplitude, compared to the usual ±80μV amplitude of the background activity. "Bad channels" can be also recovered by interpolation using the surrounding electrodes (more efficient when dense electrode arrays are available). More sophisticated methods include filtering which is now widely available on any EEG reviewing software. Eye blinks are often present during EEG experiments and they can be removed using the independent component analysis (ICA) method performed manually or automatically [4]. Recording the electrooculography (EOG) signal simultaneously with the EEGs could help to precisely and automatically remove the eye blinks. In this case, adaptive filtering proved to be relatively efficient [15]. The muscle artifacts can also severely corrupt the EEG signals. They are more difficult to remove due to the overlap with the EEG frequency band. Several studies on simulated and real data showed that the use of the blind source separation methods such as the canonical correlation analysis is a very powerful tool to remove the muscle artifacts [16, 17].

## B. Reconstruction of EEG sources

To localize brain sources and reconstruct their time-courses the following data are required: i) the scalp-recorded EEG signals, ii) the 3D positions of electrodes positioned on the head, iii) the head model which contains information about the electrical and geometrical characteristics of the head and iv) the source model which provides information about the location/orientation of dipole sources to be estimated. A template file for the 3D electrode locations is often available with the acquisition systems. However, in a patient-specific or subject-specific context, actual position may be required. A number of 3D digitizing devices allow for the registration of the electrode positions on the head (such as Fastrak Digitizer, Polhemus Inc.; Geodesic Photogrammetry System, EGI Inc.). Realistic head models employing the Boundary Element Method (BEM, surfacic case) or the Finite Element Method (FEM, volumic case) allow for accurate calculation of the electrical fields in the brain. Compared to simple spherical head models, improved realism in the description of the head geometry and tissues with associated conductivities increases the quality of the EEG forward/inverse solution. The source model is computed from the segmentation of the anatomical MRI (template or subject-specific). Usually, the white matter / grey matter interface is chosen as the source space for neocortical sources which mostly contribute to EEG. The MRI anatomy and channel locations are co-registered using the same anatomical landmarks (left and right pre-auricular points and nasion). In the following, we complement the above qualitative description of EEG source reconstruction with more formal aspects.

According to the dipole theory, EEG signals **X(t)** recorded from **M** channels can be considered as linear combinations of **P** time-varying current dipole sources **S**(t):

$$X(t) = \begin{pmatrix} x_1(t) \\ ... \\ x_M(t) \end{pmatrix} = G. \begin{pmatrix} s_1(t) \\ ... \\ s_P(t) \end{pmatrix} + N(t) = G.S(t) + N(t)$$

where **G** (M × P) is called the lead field matrix and **N**(t) is the noise. **G** reflects the contribution of each brain source to the sensors [9]. It is computed from a head model (volume conductor) and from the position of electrodes. In the case where the source distribution is constrained to a field of current dipoles homogeneously distributed over the cortex and normal to the cortical surface, the position and the orientation of the sources are defined. In the case of the methods described below, the EEG inverse problem consists of estimating the source magnitude of $\hat{S}(t) = W.X(t)$ (Eq. 1). Several algorithms have been proposed to solve this problem and estimate W based on different assumptions related to the spatiotemporal properties of sources and regularization constraints, see [9] for review. Here, we describe two methods widely-used in EEG source connectivity analysis respectively based on a minimum norm estimate and on a beamformer filter.

The *weighted Minimum Norm Estimates (wMNE)* is one of the most popular approaches. Here, **W** is estimated in such a way to produce the source distribution with the minimum power that fits the measurements in a least-square-error:

$$W_{wMNE} = BG^T(GBG^T + \lambda C)^{-1}$$

where $\lambda$ is the regularization parameter and C represents the noise covariance matrix. The wMNE algorithm compensates for the tendency of **Minimum Norm Estimate** (MNE) to favor weak and surface sources [18]. Matrix B adjusts the properties of the solution by reducing the bias inherent to the standard MNE solution. Classically, B is a diagonal matrix built from matrix G with non-zero terms inversely proportional to the norm of the lead field vectors. Note that B=I in the case where weighting is null. Practically, $\lambda$ is computed based on the Signal to Noise Ratio (SNR): $\lambda = \dfrac{1}{\text{SNR}}$.

The SNR depends on the data type. For instance, in the task-related paradigm, the pre-stimuli are usually considered as noise and the post-stimuli as the useful signal. SNR can be computed from the ratio of the signal variance over these two periods. In addition, the pre-stimuli period can be also used to compute the noise covariance matrix C. In resting state data, the computation is more difficult as the difference between the signal and baseline is very low. A long EEG segment is traditionally used to estimate the C matrix. When the noise can be assumed as spatially uniform across all channel sites then **C**=I.

Another popular inverse solution is the **beamforming**. The beamformer filter extracts the components of a signal with some specific spatial features. More specifically, it allows for scanning each source location and for retaining a signal contribution that originates from that spatial location while it rejects any contribution stemming from other locations. The weights in matrix **W** (which correspond to each specific source location) are therefore estimated one by one from the data. The data covariance matrix C is used for this purpose. One of the most widely-used beamformer is the linearly constrained minimum variance (LCMV) [19] that makes use of the following weight estimation for the source placed at a given location:

$$\mathbf{W}_{\text{beamformer}} = \left[ (\mathbf{G}^T . \mathbf{C}^{-1}) . \mathbf{G} \right]^{-1} . (\mathbf{G}^T . \mathbf{C}^{-1})$$

The two above-described methods belong to a wide set of signal processing methods aimed at solving the EEG inverse problem, i.e. estimating matrix **W** from which the dynamics of the brain sources can be reconstructed using Eq. 1. This estimation is usually done on high-resolution surface mesh (8000 or 15000 vertex for instance). However, this number of reconstructed sources is too high to perform the second step of connectivity analysis. Therefore, in practice, spatially-closed brain sources are clustered based on a set of **R** pre-defined regions of interest (ROIs), with **R** chosen as small with respect to the number of estimated sources.

To define ROIs, many anatomical or/and functional atlases are available such as the Desikan Killiany atlas (68 ROIs used in the illustrative example, figure 1B) and the Destrieux atlas (148 ROIs). This procedure leads to **R** regional time series **R(t)** each one representing the average brain activity generated by one of the **R** pre-defined brain regions. It is worth noting that 3D surface of neocortical patches is folded. In order to avoid activity cancelation due to opposite direction of dipole sources, the averaging is performed on the absolute value of the dipole moments. Averaging the time series across regions of interest (ROIs) is a simple way to produce a single time-series representative of the activity of a given extended brain source (ROI). Note that the absolute value transformation is a bit anecdotal. It accounts for calculation errors which affect a very small number of sources that are flipped to the dominant (and correct) direction before averaging. Nevertheless, there exist certainly some other approaches to estimate the activity associated with a given ROI such as the use of dimensionality reduction technique such as the Principal Component Analysis for instance.

## C. Functional and effective connectivity

Once the **R(t)** time series are reconstructed, the statistical couplings between these regional time series can be estimated. When the estimated quantity is only related to the degree of coupling, then the method is referred to as *functional connectivity*. When the objective is to estimate directionality in this coupling or causality between considered time-series, then the method is referred to as *effective connectivity*. Both functional and effective connectivity methods have been the topic of intensive research over the two past decades and many metrics are now available (review in [21]).

Concerning the functional connectivity, the most widely used methods in the EEG context are those based on the linear/nonlinear correlation, the coherence function, the phase synchronization, the mutual information and the amplitude envelope correlation (see [22] for review and [23] for comparative studies). A key issue is the performance that generally whatever the context (cognitive research or clinical application), each method has its own advantages and limitations and there is no consensus about one standard method that would outperform the other methods. In this section, we present three main families of methods: the linear correlation, the phase synchronization and the amplitude envelope correlation, as they represent the most used methods in the context of EEG source connectivity.

*The cross-correlation coefficient ($r_{xy}^2$)* is one of the oldest and probably the most classical measure of interdependence between two time series. Conceptually very close to the so-called Pearson's correlation coefficient in statistics, it is a measure of the linear correlation between two, signals *x* and *y,* possibly delayed by $\tau$:

$$r_{xy}^2(\tau) = \frac{\text{cov}^2(x(t), y(t+\tau))}{(\sigma_{x(t)}\sigma_{y(t+\tau)})^2} \quad \text{(Eq. 2)}$$

where $\sigma$ and cov denote the standard deviation and the covariance, respectively. Starting from Eq. 2, the metrics $r_{xy}^2$ classically used to characterize the coupling between *x* and *y* is given by:

$$r_{xy}^2 = \max[r_{xy}^2(\tau)]$$
$$-\tau_{\max} < \tau < \tau_{\max}$$

where $\tau_{\max}$ denotes the maximum time shift between the two signals.

The second family of method is the *Phase synchronization (PS)*. It is well known that the phases of two time series can be synchronized even if their amplitudes are independent. The general principle of the *PS* is to detect the presence of a phase locking between two systems defined as:

$$\varphi(t) = |\Phi_x(t) - \Phi_y(t)| \leq C$$

where $\Phi_x(t)$, $\Phi_y(t)$ are the unwrapped phases of the signals *x* and *y* at time bins *t* and *C* is a constant. The first step is to extract the instantaneous phase of each signal. Two different techniques can be used: the Hilbert transform and the wavelet transform. It was shown that the application of both approaches produces relatively close results. The second step is the definition of a metric that measures the synchronization degree between the estimated phases. Several measures have been proposed to measure PS between two signals.

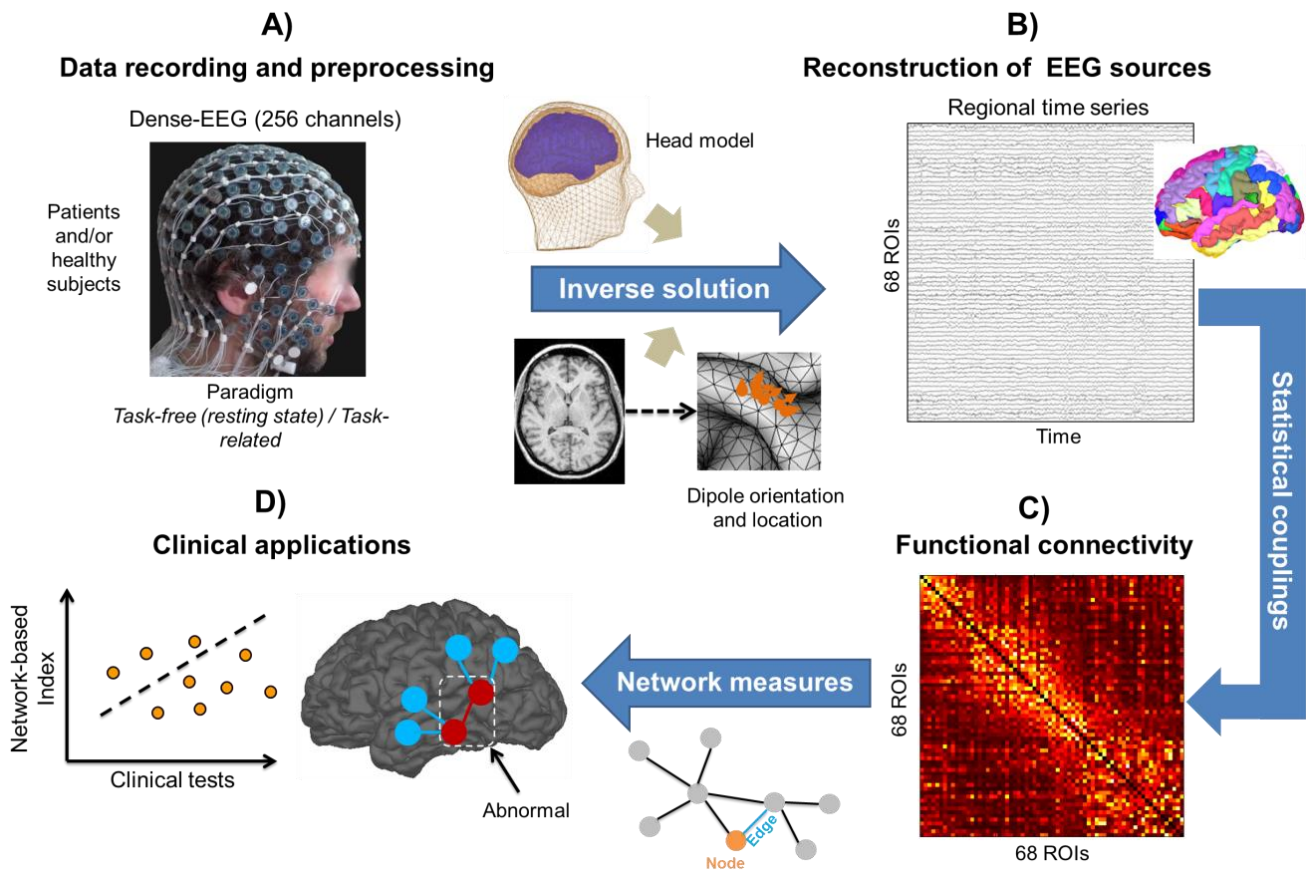

**Fig. 2.** *From EEG recording and pre-processing to brain networks and applications*. **A)** EEG data can be recorded during task-related (evoked response) or task-free (resting state) paradigms. These recordings can be performed on patients as well on healthy subjects. Signals are pre-processed using artifact removal and filtering techniques, for instance. Resulting signals constitute the input to the source connectivity method. **B)** To reconstruct EEG sources, the lead field matrix (contribution of each cortical source to the scalp sensors) is required. It is computed from i) a multiple layer head model (volume conductor) is obtained from MRI segmentation and ii) the position of scalp electrodes. The Boundary Element Method (BEM), illustrated here, is one of the available numerical methods. It is classically used in the case of realistic multiple layer head models (skin, skull, CSF, grey matter, white matter). Using segmented MRI data, the source distribution is constrained to a field of current dipoles homogeneously distributed over the cortex and normal to the cortical surface. The dynamics of the reconstructed sources are then estimated by solving the inverse problem which consists of estimating the remaining free parameter, i.e. the moment of the dipoles. A source space with defined regions of interest (ROIs) is usually used given a number of regional time series (68 ROIs extracted from a Desikan atlas [20] in this illustrative example), **C)** Once the regional time series are reconstructed, the functional connectivity can be then estimated by computing the statistical couplings between these time series. This produces an adjacency matrix which represents the pair-wise functional connections between all the ROIs, **D)** Once nodes and edges have been defined, network topological properties (organization) can be studied by graph-theory based analysis. These quantitative metrics can be used for cognitive research or in a clinical perspective such as the localization of abnormal epileptic networks or the computation of biomarkers of cognitive decline in neurodegenerative diseases.

The *phase locking value* (PLV) [24] is defined as

$$\text{PLV} = \left| \langle e^{i\varphi(t)} \rangle \right|$$

where $\langle . \rangle$ denotes average over time and trials.

The *phase-lag index* (PLI) [25], quantifies the asymmetry of the phase difference, rendering it insensitive to shared signals at zero phase lag

$$\text{PLI} = |\langle \text{sign}\varphi(t) \rangle|$$

Another method used in MEG/EEG source connectivity is the *amplitude envelope correlation (AEC)*. This method consists of estimating the amplitude correlation between signals using linear correlations (or partial correlations) of the envelopes of 'filtered' signals. The envelopes of the signals can be computed using Hilbert transform [26].

The $r^2$, PLV, PLI and AEC values range from 0 (independent signals) to 1 (fully correlated/synchronized signals).

The above described *functional connectivity* methods only consider the degree of coupling. In contrast, *effective connectivity* methods are aimed to estimate the causality (in the sense of Granger) or the directionality of coupling between the signals. Several methods have been proposed based on multivariate autoregressive model (MVAR), such as the directed transfer function (DTF) and the partial directed coherence (PDC) [27].

As an example, we describe here the method based on the parametric representation of multichannel time series, which is widely used to study causal brain interactions. For signals **X**(t) with **M** dimensions, the multivariate autoregressive (MVAR) with order $p$ can be defined as:

$$X(t) = \sum_{i=1}^{p} A(i) X(t-i) + \varepsilon(t)$$

where $\varepsilon(t)$ denotes the additive noise and $A(i)$ are the model coefficients ($M \times M$). This time domain representation can be transformed into frequency domain.

$$X(f) = A^{-1}(f)\varepsilon(f) = H(f)\varepsilon(f)$$

where $H(f)$ is the transfer function and $A(f)$ is the Fourier transform of the coefficients. Using MVAR coefficients, the PDC estimator, characterizes the outflow from channel $j$ to channel $i$ at frequency $f$, is defined as:

$$PDC_{ij}^{2} = \frac{A_{ij}^{2}(f)}{\sum_{r=1}^{k} A_{rj}^{2}(f)}$$

and the DTF estimator, which describes the causal influence of channel $j$ on channel $i$ at frequency $f$, is defined as:

$$DTF_{ij}^{2}(f) = \frac{|H_{ij}(f)|^2}{\sum_{r=1}^{k} |H_{ir}(f)|^2}$$

Other methods are also available to compute effective connectivity. They are based on a directionality index derived from nonlinear regression analysis [28], on the transfer entropy [29] or on the combination of effective connectivity with neural mass models identified from time series. This latter method is known as dynamic causal modeling (DCM) [30]. For the sake of space, they are not described here. Readers may refer to [21] for review.

## D. Network measures

From the previous step, and whatever the connectivity method being used (functional or effective), an **R×R** adjacency matrix is produced. This matrix represents the pair-wise connections between all the ROIs. An example of functional connectivity matrix for R=68 is presented in figure 1C. In order to retain significant interactions, these matrices are usually thresholded (keep only the top 10% of connections for instance) to distinguish real functional connections from spurious connections. A variety of thresholding methods are available, but no method is free of bias. It is then cautious to perform studies across different values of thresholds (in addition to the use of alternative strategies) to ensure that the obtained findings are robust to this methodological factor. In the context of EEG, other techniques are also available to test the significance of interactions such as the use of surrogate data analysis. Readers can check [31] for a complete overview about most network-related methodological issues.

Interestingly, this **R×R** adjacency matrix can be characterized and quantified using network measures derived from graph theory. Graph theory is a branch of mathematics focused on the analysis of systems consisting of interconnected elements. Such a system can be represented as a graph in which nodes (or vertices) are connected by edges (or links). In the context of brain networks, the nodes represent the brain regions and the edges reflect the functional and/or effective connections. Once nodes and edges are defined, network topological properties can be studied by graph-theory metrics. As illustrated in figure 1, these quantitative metrics can be used to characterize the normal brain network architecture during rest or during cognitive functions. They can also be used in a clinical perspective such as the localization of epileptic zones (figure 1D, right) or the development of neuromarkers for other brain disorders (figure 1D, left).

A simple graph can be represented by $G = (V, E)$ where $V$ is the set of nodes and $E$ is the set of edges. In the weighted undirected graph, each node can be identified by integer value $i = 1, 2, ..., N$ and an edge can be identified by $(i, j)$ represents the connection going from node $i$ to node $j$ to which a weight $A_{ij}$ can be associated. Some of the main network measures, illustrated in Figure 3, are briefly described hereafter.

The *Degree (d)* denotes the total number of links connected to a given node (figure 3A).

The *clustering coefficient (C)* reflects the tendency of a network to form topologically local circuits (figure 3B). For a given node $i$ with degree $d$, the local $C_i$ is defined as:

$$C_i = \frac{2L_i}{d_i(d_i - 1)}$$

where $L_i$ denotes the number of links between the $d_i$ neighbours of node $i$. $C_i$ varies between 0 and 1 and it is considered as the main graph metric of information segregation in networks. The more the neighborhood of node $i$ are densely interconnected, the higher is its local clustering coefficient. This network measure will be used in real application described below (figure 4).

The *Path length* is the average weighted shortest path length often used as a measure for global integration of the network. It is defined as the harmonic average of the shortest paths between all possible vertexes pairs in the network, where the shortest path between two vertices is defined as the path with the largest total weight. The *global efficiency ($E_G$)* of a network is the inverse of the characteristic path length. Several studies have used $E_G$ as a measure of information processing capability. The global efficiency is a measure of integrated and parallel information-processing (figure 3C).

Generally speaking, in a network, the functional value of a node is proportional to the number of paths in which it participates. A way to find the critical nodes in a brain network is to calculate the betweenness centrality of each node. The *Betweenness Centrality* (BC) of a node is defined as the number of shortest paths in the network that pass through the node normalized by the total number of shortest paths. Another metric used to characterize the network topology is *Modularity* which denotes the partitioning of the associated graph into a number of clusters or modules (also called communities). A network module is defined as a subset of nodes in the graph that are more densely connected to other nodes within the same module than the nodes in the other modules. *Hubness* can be measured based on the intra-modular connectivity ($Z$) and participation coefficient ($PC$). Once the modularity is calculated and optimal modules have been identified, the $Z$ and the $PC$ metrics are computed for each node. The nodes are classified as hubs if their $Z$ is higher than a defined threshold $T$, otherwise they are classified as non-hubs.

Using $PC$, a hub can be classified as a *provincial hub* where the nodes are mostly connected to nodes within its own module, or as a *connector hub* where the node have diverse connectivity across several different modules in the network (figure 3D). For dynamic networks, the modularity can be computed using the multislice network modularity algorithm [32].

*E. Software*

Several software packages were developed to process EEG signals such as EEGLAB, CARTOOL, Fieldtrip and Brainstorm. In addition, a number of toolboxes have been proposed to analyze and visualize complex networks such as Brain Connectivity Toolbox (BCT), BrainNet Viewer, the GCCA toolbox, the connectome mapper, Gephi, the connectome Viewer, the eConnectome, EEGNET, the Connectome Visualization Utility (CVU) and GraphVar. However, a comprehensive tool that implements the complete pipeline from EEG processing to analysis and visualization of brain networks is still missing. Table 1 provides a list of matlab-based tools along with main functionalities in term of EEG pre-processing and inverse solution, functional and effective connectivity measures and network characterization and visualization.

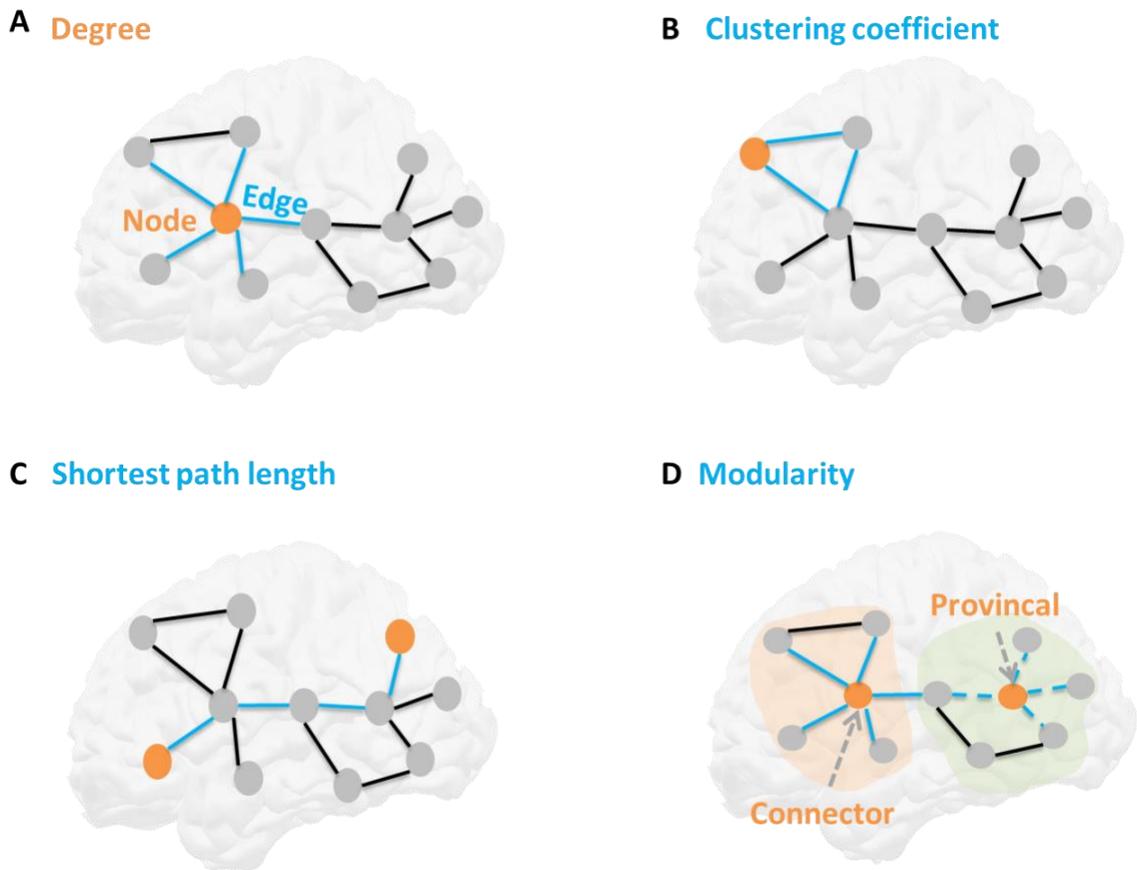

**Fig. 3. Some metrics used in undirected graphs built from connectivity methods applied to brain sources reconstructed from EEG. A. A node with high degree compared to other nodes. B. The clustering coefficient of a node is computed as the number of triangles attached to a node, relative to the total possible number of triangles. C. The shortest path between two nodes. The characteristic path length of a network is the average path length between every pair of nodes. D. Illustration of a modular decomposition of the network. Two modules have been identified, as represented by the different background colors. Nodes within a module are strongly connected with each other and sparsely connected with nodes in other modules. Such decomposition allows for analysis of node roles and hub category: a *provincial* hub is highly connected within its own module while a *connector* hub has connections distributed across modules.**

IV. DYNAMIC RECONFIGURATION OF FUNCTIONAL BRAIN NETWORKS

Tracking the temporal dynamics of brain networks is an issue of great interest in cognition and neuropathology. The term "brain network reconfiguration" refers to slow changes across the lifetime due to experience as well as to rapid spontaneous or evoked changes in response to external stimuli or perturbations [33]. For instance, an important challenge is to temporally follow, over very short time duration (sub-second), changes in functional brain networks involved in a cognitive task. A key advantage of EEG is its excellent temporal resolution that offers the unique opportunity to track large-scale brain networks over time. This excellent temporal resolution permits the analysis of the dynamic properties of brain processes, an issue so far addressed in a few studies dealing with cognitive activity or with resting state (participants are not involved in a particular task).

Many studies have reported that a number of brain regions remain highly functionally connected even when subjects are at rest (closed or open eyes). In this context, EEG allows for tracking of the temporal dynamics of resting state networks (RSNs) at sub-second time scale, a result that is not reachable using functional magnetic resonance imaging (fMRI). In these studies, results showed the key role of some specific brain regions such as the posterior cingulate cortex and the prefrontal cortex (forming the so-called Default Mode Networks -DMN-), in maintaining efficient temporal communication in the whole brain. Other studies focused on assessing the temporal transitions between the main resting states networks such as the visual network, the audio network and the dorsal attentional network [4]. Recently, the EEG source connectivity method was also used to track the task-related networks, i.e. following the trajectory of the information processing in the human brain from the beginning to the end of a short duration cognitive task (sub-second). The brain network reconfiguration was tracked during visual, motor and memory tasks [3, 34, 35]. Using clustering algorithms (such as k-means algorithm), these studies showed that any cognitive function can be decomposed into a set of *brain network states (BNS)* that reflect the underlying cognitive processes (visual/semantic processing and access to memory for instance).

TABLE 1

MATLAB-BASED TOOLBOXES FOR PREPROCESSING, SOURCE ESTIMATION, FUNCTIONAL/EFFECTIVE CONNECTIVITY MEASURES AND NETWORK ANALYSIS/VISUALIZATION

| Software | Web page | EEG Pre-processing | EEG Inverse solution | FC/EC | Network measures | Network visualization |
| --- | --- | --- | --- | --- | --- | --- |
| Brainstorm | http://neuroimage.usc.edu/brainstorm/ | ☺ | ☺ | ☺ | | |
| EEGLAB | https://sccn.ucsd.edu/eeglab/ | ☺ | ☺ | | | |
| FieldTrip | http://www.fieldtriptoolbox.org/ | ☺ | ☺ | ☺ | | |
| eConnectome | http://econnectome.umn.edu/ | ☺ | ☺ | ☺ | | ☺ |
| EEGNET | https://sites.google.com/site/eegnetworks/ | | ☺ | ☺ | ☺ | ☺ |
| Conn | https://www.nitrc.org/projects/conn/ | | | ☺ | ☺ | ☺ |
| FCT | https://sites.google.com/site/functionalconnectivitytoolbox/ | | | | ☺ | |
| BCT | https://sites.google.com/site/bctnet/ | | | | ☺ | ☺ |
| BNV | https://www.nitrc.org/projects/bnv | | | | ☺ | ☺ |
| GraphVar | https://www.nitrc.org/projects/graphvar/ | | | | ☺ | ☺ |
| NBS | https://www.nitrc.org/projects/nbs/ | | | | ☺ | ☺ |

In figure 4, we report some novel results showing the performance of the EEG source connectivity method in the context of a visual cognitive task. Two categories of visual stimuli were presented on a screen: meaningful (animal, tools…) and meaningless (scrambled) images. Participants (N=20) were asked to name the presented visual stimuli. By using the combination of the wMNE and PLV (see section ***From EEG signals to cortical network B, C***), computed over trials (n=120), functional networks in the EEG gamma band (30-45 Hz) were obtained. A *k*-means clustering algorithm was then applied to segment the EEG responses and led to identify four *BNS*. Details about this segmentation algorithm developed in the context the picture naming task can be found in [3]. We were then interested in the difference in term of network topology between both categories. We computed the clustering coefficient reflecting the local information processing, as defined above, for each brain region (R=68, Desikan-Killiany atlas) and we retained regions leading to significant differences between both conditions (t-test, *p*<0.01, corrected for multiple comparisons using False Discovery Rate -FDR- method). Figure 4 shows clearly that the clustering coefficient of the network is not the same for both categories over time, suggesting that the process of information segregation in the brain is different in the two conditions (meaningful vs.

meaningless). Interestingly, the method also showed the implication of the temporal lobe for all states, which is widely reported to be related to the semantic processing in the brain.

## V. EEG SOURCE CONNECTIVITY IN BRAIN DISORDERS

Converging evidence suggests that perturbations in the brain are rarely limited to a single region. As the brain is a complex network of structurally and functionally interconnected regions, local dysfunctions often propagate and affect other regions, resulting in large-scale network alterations [2]. These dysfunctions likely occur at both axonal and synaptic level. A typical example is the rapid spread of partial (focal) epileptic activity into at the onset of seizures which rapidly involves spatially distributed brain regions [36]. Along the same line, but on a longer time scale, it is now established that the progressive evolution in Alzheimer's Disease (AD) and Parkinson's Disease (PD) is also related to pathological changes in large-scale networks, although these neurodegenerative diseases have a focal onset [37].

Therefore, from a clinical perspective, the demand is high for non-invasive and easy-to-use methods to identify pathological networks like those involved in epilepsy. In addition, the demand is also high for novel 'neuromarkers' in other neurological diseases able to characterize network alterations and associated cognitive deficits in PD and AD patients, in particular at early stage. In this context, EEG has some major assets since it is a non-invasive, easy to use and clinically available technique. Therefore, and as shown by our recent studies [11, 38], EEG source connectivity methods could provide some responses to clinical demand, provided that appropriate information processing is performed. Below we describe the main applications of EEG source connectivity in neurological disorders.

### A. Epilepsy

Ding et al. [40] were the first to apply functional connectivity to EEG source signals in epileptic patients. Authors showed that the method was able to distinguish the primary sources responsible for the seizure generation from the secondary sources involved in the seizure propagation. In a follow-up study, Lu et al. [41] applied the method to EEG recordings (76 channels) performed in patients with partial epilepsy. Authors found that EEG source connectivity method leads to correct seizure onset localization, as compared to invasive recordings. They also report the need of high number of electrodes to better estimate the epileptic network.

Combined with graph theory based analysis, Vecchio et al. [42] used EEG source connectivity in patients suffering from fronto-temporal epilepsy. Authors reported a significant increase of the local (characterized by the clustering coefficient) and global (computed using the characteristic path length) connectivity in the alpha band in the ipsilateral hemisphere as compared with the contra-lateral hemisphere. Applied to 16 patients, Coito et al. [43] investigated the directionality of the interactions between brain regions during interictal spikes estimated with effective connectivity methods applied at source level. In addition to good matching with invasive recordings, authors showed also a relationship between the connectivity patterns and the neuropsychological results obtained in patients.

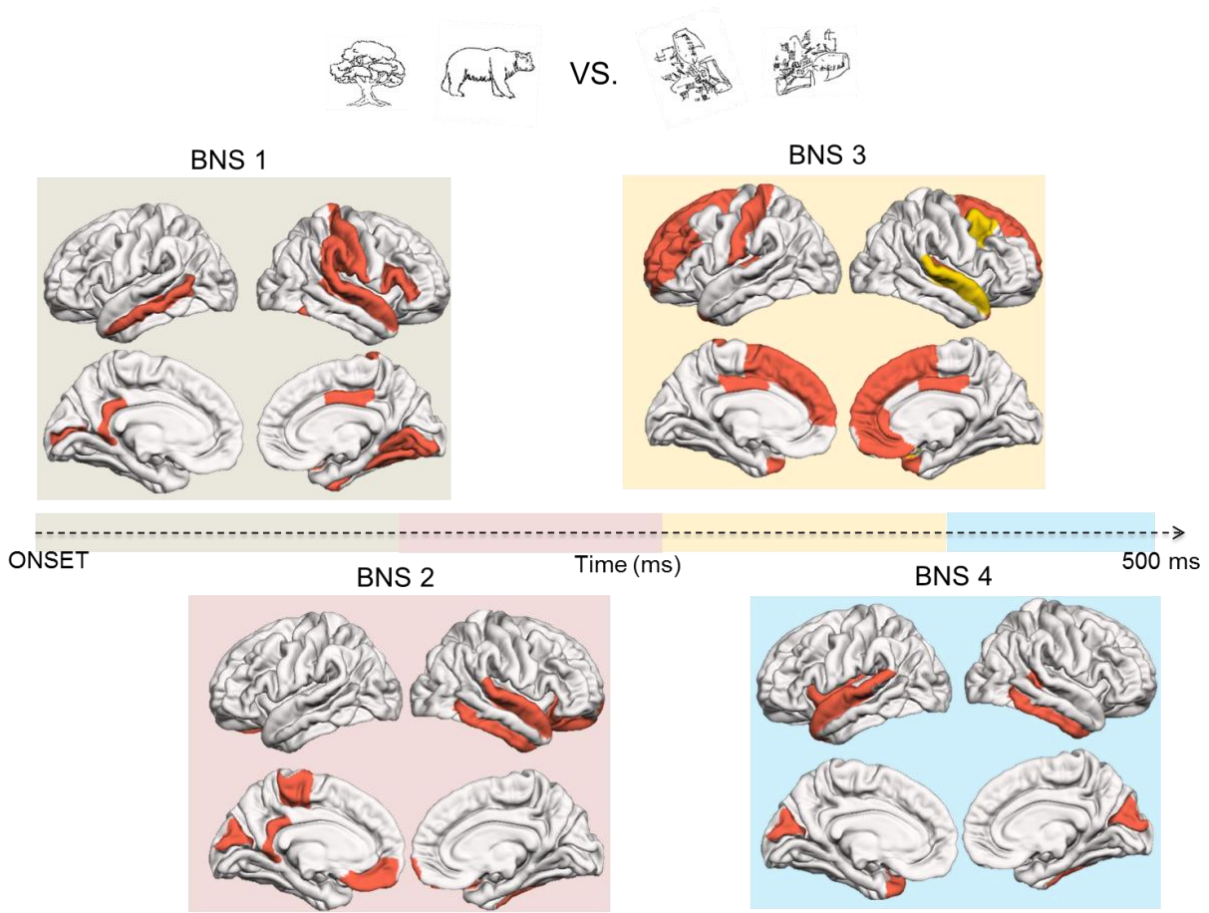

**Fig. 4: Application of EEG source connectivity to tracking dynamic reshaping of functional brain networks during visual objects recogniton. Brain regions showing significant difference ($p<0.05$) in term of clustering coefficient over time periods (obtained using a k-means clustering algorithm described in [39]) between the two categories of objets: meaningful (left) and meaningless (right). Orange red: uncorrected for multiple comparisons, gold: corrected for multiple comparisons using FDR approach. BNS: Brain Network State. These results show the power of the EEG source connectivity method to track very short cogitive task (<1s) and to reveal brain regions that associate the meaning to visual objects recognized in the human brain.**

Here, we present a model-based evaluation of EEG source connectivity methods aimed at identifying epileptogenic networks from scalp recordings (figure 5). We performed a joint comparison of two inverse solutions algorithms (wMNE and dSPM) and two connectivity measures (PLV and $r^2$) using data simulated from a biophysical/physiological model that allows for the generation, at cortical level, of realistic interictal epileptic spikes that also reflect in scalp EEG signals. We used a network-based similarity index to compare the network identified by each of the inverse/connectivity combination with the original network simulated in the model. The main advantage of this algorithm, called SimiNet, is that it takes into consideration the physical locations of the nodes to compute the network-similarity, which is a crucial element when dealing with brain networks. The nodes showed in figure 6B represents the physical locations of the generated sources while in figure 6A represents nodes with the highest 5% strength values (most important nodes in the network). Edges are not shown to enhance visualization.

Globally, results revealed that the choice of the inverse/connectivity combination can have a significant impact on the networks identified from scalp EEG signals, (figure 6A). They also showed that methods based on phase synchronisation (PLV) combined with the wMNE inverse algorithm show higher performance in term of similarity between reference network and identified network, as compared with other combinations (figure 6C). Other methods and other network scenarios were tested in [38]. Interestingly, the same combination exhibited the highest performance. Finally, it worth noting that applying this combination on real dense EEG (256 channels) data recorded in epileptic patients candidate to surgery showed a very good matching between scalp-EEG-based networks and intracerebral-EEG-based networks, as reported in [38].

## B. *Neurodegenerative diseases*

Neurodegenerative diseases are associated with distinct patterns of functional network dysfunction [37]. The main motivation of using EEG source connectivity here is to found an association between the degree of cognitive deficits, on one hand, and the alterations in the functional brain networks, on the other hand. The hypothesis is that cognitive impairment gradually worsens with the progressive alteration of brain functional connectivity. Beside Neurodegenerative diseases, EEG source connectivity was also used in other applications such as schizophrenia [44], major depression [45], pain [46] and obsessive compulsive disorder [47]. In this section, we highlight some recent results obtained with this approach in Parkinson's disease (PD).

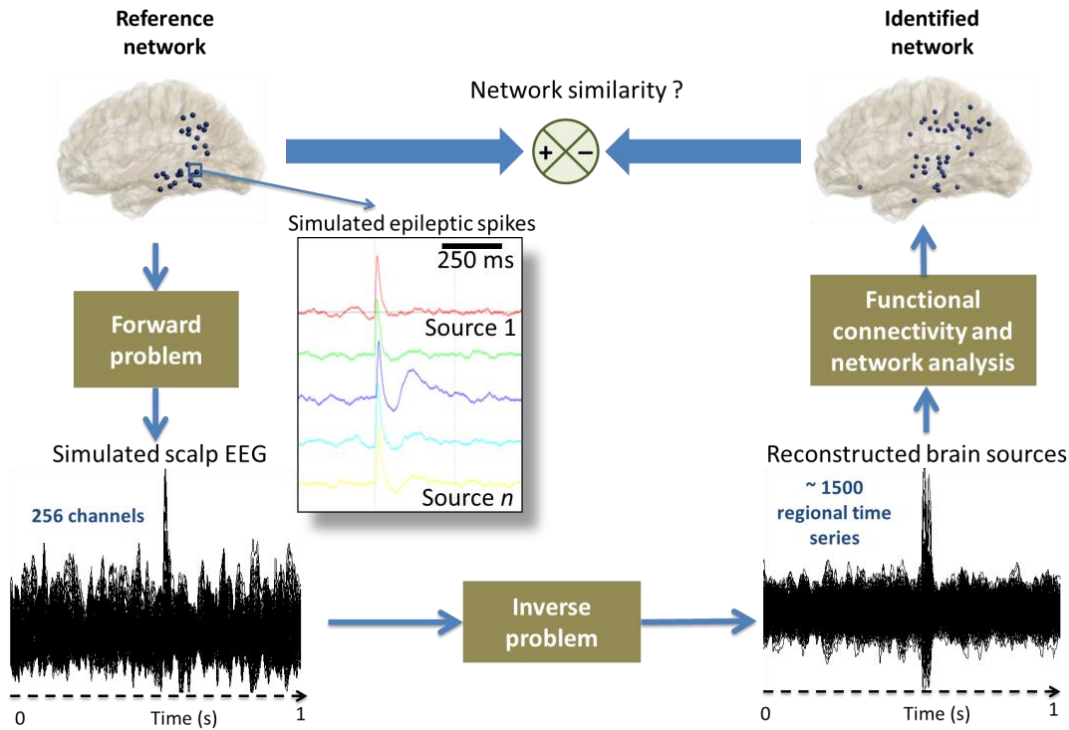

**Fig. 5. Model-based evaluation of EEG source connectivity methods aimed at identifying epileptogenic networks from scalp recordings. First, a spatially-distributed epileptogenic network is generated by a physiological model (coupled neural masses generating epileptic spikes). This network is considered as the 'ground truth'. By solving the forward problem, synthetic dense EEG data are generated. These simulated signals are then used evaluate the performance of EEG source connectivity methods according to their ability to recover the 'reference' network. Different combinations of methods where used to solve the inverse problem and reconstruct the dynamics of cortical sources. For each combination, the identified network is compared with the original network using a 'similarity index' accounting for topological features (3D position of nodes and edges) of matched networks.**

Dense EEG (122 channels) source connectivity was used by Herz et al. [48] in patients with PD. Results revealed the effect of dopamine in the reconfiguration of the prefrontal-premotor connectivity. Using MEG source connectivity, decreases in alpha1 [8-10Hz] and alpha2 [10-13Hz] frequency band connectivity were observed in PD patients. Most of the alterations were located in temporal regions. In a 4-year longitudinal study, MEG source connectivity was also applied, by the same team, on 70 PD patients to tracking the resting state of networks in the aim of assessing the possible follow up of the disease progression [49]. Authors reported a progressive decrease in the local clustering network measure in multiple frequency bands together with a decrease in path lengths at the alpha2 frequency band. These alterations were related to a worsening in the motor function and cognitive performance. This study was the first to show that network measures (such as the local/global efficiency) may lead to promising neuromarkers of PD progression. Using dense-EEG from 124 PD patients, we recently reported progressive disruptions in functional connectivity between three patient groups: cognitively intact patients, patients with mild cognitive deficits and patients with severe cognitive deficits. Our findings indicate that functional connectivity decreases with the worsening of cognitive performance, suggesting that it can potentially be used to devise novel neuromarkers of cognitive impairment in PD patients [11].

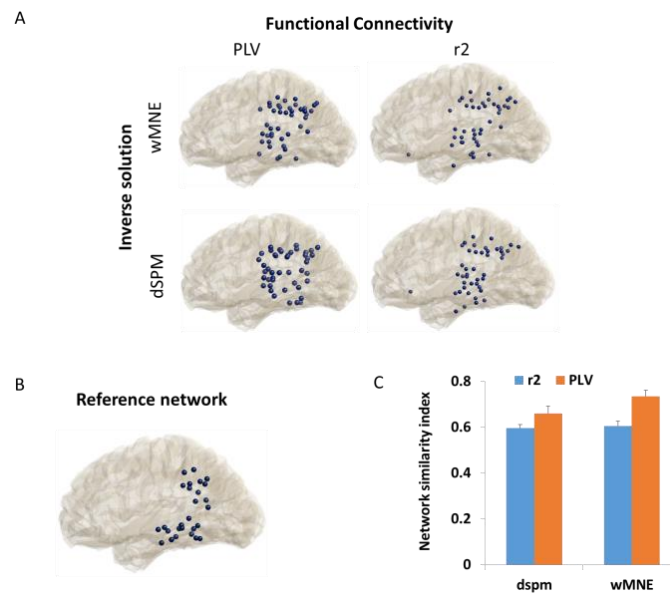

Fig. 6. A) Brain networks obtained using two different inverse (wMNE and dSPM) and functional connectivity ($r^2$ and PLV) methods. B) The original network (ground truth), C) Values (mean ± standard deviation) of the similarity index computed between the network identified for each combination and the reference 'epileptogenic' network used to simulate dense-EEG data. Adapted from [38].

VI. DISCUSSION

Electroencephalography (EEG) consists in measuring the brain electrical activity using electrodes positioned on the scalp. A key feature of EEG is its intrinsically excellent time resolution that makes it unique for tracking the fast reconfiguration of functional networks of neuronal assemblies distributed in the cerebral cortex. Emerging evidence shows that the functional brain connectivity computed at scalp level ("electrode space") does not allow for relevant interpretation of anatomically interacting areas as estimates are severely corrupted by the volume conduction effects, see [6, 7] for recent comments. One more efficient solution, described in this review, is to compute the functional connectivity at the level of the brain sources ("source space"). This method, named EEG source connectivity, combines the excellent resolution of EEG and adds a very good to excellent spatial resolution depending on the granularity (coarse to fine grain) of the source model that is used to solve the EEG inverse problem and subsequently identify networks at cortical level.

*Spatial leakage*:

A critical issue often raised in the computation of connectivity at source level is the 'spatial leakage'. Indeed, as source estimates are spatially correlated, a leakage of inferred sources into their local neighborhood often occurs. When the connectivity method ignores this effect, "false" connectivity values computed between distant sources may be interpreted as function connectivity although they reflect the fact that sources share components of the same sensor signal. To address this issue, several strategies have been proposed to remove zero-lag correlations before performing connectivity analyses. Other studies suggest that only the long-range connections should be kept. However, these solutions may suppress important correlations that might happen at zero-lag or even between close regions. As EEG source connectivity is still a relatively new field compared to fMRI connectivity analysis, more methodological efforts are still needed to completely overcome issues such as mixing and spatial leakage. We also advice the use of multimodal recordings such as EEG/fMRI, which can benefit from the excellent spatial resolution of the fMRI and the excellent time resolution of the EEG and can help to cross-validate the results from both techniques.

*Consistency of inverse/connectivity measures:* Although all reported EEG studies include two main steps (EEG inverse problem followed by source connectivity estimation), they differ from a methodological perspective. Indeed, different algorithms were used to reconstruct cortical sources. In these algorithms, various mathematical assumptions are used for the regularisation of an ill-posed inverse problem. Main assumptions relate to sources with minimum energy, time/space sparsity and possible correlation between the reconstructed sources. A plethora of functional and effective connectivity measures were also proposed to measure statistical couplings between regional time series. Therefore, the natural question that is raised is what combination of inverse/connectivity method should be used to enhance the global performance and to guarantee the relevance of results in term of identified brain networks? Unfortunately, there is no answer to this question. As each of the inverse and functional/effective methods has its own assumptions and characteristics, there is no consensus yet about the best combination. This crucial issue has been addressed in a number of studies showing that selected methods (inverse solution and connectivity measure) directly impacts the topological/statistical properties of networks identified from EEG surface signals. Recently, Mahjoory et al. have evaluated the effect of the anatomical templates, head models, inverse solutions and the software implementations. Authors showed variability between the inverse solution algorithms (mainly LCMV and wMNE). Also, the functional connectivity measures were much more consistent across the variables as compared with measures obtained with effective methods.

We have also conducted two comparative studies regarding the choice of the "optimal" combination on inverse/connectivity method. In both studies, our intent was to maximize the *a priori* information ("ground-truth") about the brain networks which were supposed to be identified from dense-EEG. In the first one [14], we used a widely-used cognitive task (picture naming) for which strong literature background was available, essentially coming from fMRI studies. In the second study [38], we used a modelling approach in which epileptogenic networks are used to simulate dense-EEG data that was subsequently used to evaluate EEG source connectivity. We then compared the network obtained by each of the inverse/connectivity combination with the reference network using a network-similarity index proposed recently in our team. Interestingly, both comparative studies led to the same conclusion: a strong variability is observed among the tested combinations but the results provided by the wMNE/PLV combination shows consistency and always exhibits the highest performance in term of matching between the estimated and the reference network. This result might be explained by the fact that wMNE relies on reasonable "physiological" assumptions (position and orientation of sources). The only "mathematical" assumption is that the solution has the lowest energy. It is worth noting that this assumption could also be interpreted physiologically in term of minimal energetic cost in the brain during task performance or at rest [50]. Regarding the second step, the PLV method estimates the phase synchronisation between EEG oscillations. Therefore, this method is in line with the concept of "communication through coherence" (CTC) in the brain in which synchronization between locally-generated signals is a crucial mechanism in brain function. In the context of EEG source connectivity, the PLV method in particular, and more generally the phase synchronization methods precisely reflect the underlying synchronization between the brain signals generated by distant sources [25]. Altogether, these features may explain the good performance of this combination of methods, in particular in the assessment of brain networks involved in cognitive activity.

*Clinical impact:* A growing body of evidence suggests that brain disorders are related to alterations in functional connections between brain regions, disrupting the normal large-scale brain network organization and function [2, 51]. A first statement from this review is that the extraction of valuable information about pathological brain networks from EEG is challenging but reachable. A second remark is that clinical practice will certainly change in the next years. Indeed, and although the combination of the EEG source connectivity with network science is still a young research field, results reported over the last few years are, clinically, very promising. It is likely that the use of novel tools allowing for characterization and quantification of the identified networks (which is the case of modern network science using graph theory based analysis) will develop and spread to clinics.

However, most of the studies reported and discussed in this review were generally performed on relatively small groups of patients. Due to the diversity of methodological approaches (candidate inverse solution algorithms and functional/effective connectivity measures, impact of the number of electrodes…) and the number of possible conditions (task-related vs. task-free paradigms), the comparison of results is still difficult. Further studies on larger cohorts of patients will certainly contribute to standardizing the analysis conditions.

In the context of epilepsy, one of the main clinical challenges is the delineation of the epileptogenic zone (EZ) electrophysiologically defined as the primary zone of organization of ictal discharges. In drug-resistant partial epilepsies, when surgery can be indicated, the resection of the EZ showed to be sufficient to significantly reduce the occurrence of seizures and even lead to seizure freedom. Yet, there exists no available technique able to precisely define the EZ. In this context, the EEG source connectivity method showed encouraging results to estimate epileptogenic networks from non-invasive recordings. In some studies, a good matching could be obtained with intracerebral invasive recordings [38, 43, 52].

In the context of neurodegenerative and psychiatric disorders, the main challenge is to develop methods that allow for establishing a relationship between i) the degree of cognitive deficits, and ii) the alterations in the functional connectivity of brain networks. To have direct clinical impact, these new methods should be non-invasive, easy to use and widely available in clinics. This is already the case of EEG technique (MEG is still more "research-oriented"). These disorders share a common feature, i.e., they are characterized by disturbances in large-scale neuronal networks. In this context, EEG source connectivity methods seem to have potential for identify dys-functional networks but also for opening new perspectives in term of neuromarkers of cognitive impairment. Results reported so far and synthetized in this article show that this objective is reachable provided that appropriate data processing is applied to sufficiently large databases [53].

*Limitations and future directions:*

Here, our intent is to review a number of recent developments based on EEG source connectivity which is considered to have a great potential for brain research. This field is not mature yet and a complete validation procedure is missing so far. However, this absence of validation is not inexorable and should not prevent us to increase our research efforts in this field. For instance, progress will certainly be made with future developments like the simultaneous recording of intracerebral and scalp EEG data that will be further used as a ground-truth to evaluate proposed algorithms, at least in patients with drug-resistant epilepsy. Some limitations and future directions are summarized hereafter:

First, EEG signals reflect a mixing of activities generated by neuronal sources arranged as assemblies. As well described by bio-electromagnetic models [19] and experimental studies [20], it is known that synaptic activation lead to the formation of a sink and a source at the level of neurons which can then be viewed as elementary current dipoles. In the case where neurons are geometrically aligned (like pyramidal cells organized "in palisade" in cortical structures) then the dipole contributions tend to sum up instead of cancelling out. These biophysical considerations explain why summed post-synaptic potentials (PSPs), either excitatory (EPSPs) or inhibitory (IPSPs) generated at the level of pyramidal cells located in the cerebral cortex are the major contribution to EEG signals recorded distantly from sources (typically with electrodes positioned on the head). These issues explain why the dipole model is the most suitable to solve the inverse problem. Nevertheless, more efforts to overcome some of the limitations of the dipole model (mainly the spatial limitations) will certainly improve the EEG forward/inverse solutions. Note that from a bio-signal processing viewpoint, the generation mechanisms of the EEG signals are considered as random (non-deterministic) processes.

Moreover, it is also noteworthy that the generation of action potentials (APs) and PSPs in networks of neurons result from complex nonlinear processes that cannot be analytically described. Consequently, local field potentials (recorded by intracerebral electrodes) and EEG signals (recorded by scalp electrodes) are random signals: they take random values at any given time, they cannot be predicted and they can only be characterized statistically. Nevertheless, at given time t, the relationship between the neuronal sources and the sensors is fully determined by biophysical factors: the position and orientation of equivalent dipoles, the source-sensor distance, the volume conductor properties (conductivity of the various layers). Typically, for EEG, the equation $X(t)=GS(t)+N(t)$ describes the relationships between the cortical sources $S(t)$ and the signals collected at scalp electrodes $X(t)$. In this equation, $S(t)$ is the random fraction of the EEG signal. G is the leadfield matrix that describes the deterministic quasi-instantaneous projection of signal sources on scalp electrodes. $N(t)$ is the "measurement noise" inherent to any acquisition procedure.

Second, volume conduction effects are prominent in the electrode space. Connectivity analysis at source level was shown to reduce the effect of volume conduction as connectivity methods are applied to "local" time-series (analogous to local field potentials) generated by cortical neuronal assemblies modelled as current dipole sources. Nevertheless, these so-called "mixing effects" can also occur in the source space but can be reduced by an appropriate choice of connectivity measures. Indeed, inverse methods are characterized by their own spatial resolution, i.e. their ability to "separate" spatially-closed sources which depends on methodological assumptions. Therefore, one should be cautious with the interpretation of brain connectivity measures even when performed at the source level since the hypothesis that part of the measured coupling is also caused the mixing of sources cannot be ruled out.

Third, every functional/effective connectivity measure has its own strengths and weaknesses. False functional couplings can be generated by some connectivity methods when applied to mixed signals such as estimated brain sources. To address this issue, a number of methods were developed based on the rejection of zero-lag correlation. In particular, "unmixing" methods, called "leakage correction", have been reported which force the reconstructed signals to have zero cross-correlation at lag zero [54]. Although handling this problem -theoretically- helps interpretation, very recent study showed that the current correction methods also produce erroneous human connectomes under very broad conditions [55].

Fourth, over the past decade, graph theory has become a well-established approach in the network neuroscience field [1]. It provides complementary information to source connectivity methods by quantifying structural, functional and/or statistical aspects of identified brain networks. This field is moving very fast and, in this respect, we stress on the need for more validation studies regarding the use of graph theory-based approaches in the context of EEG/MEG source connectivity analysis. This issue is in line with recent few attempts to evaluate other parameters involved in the EEG source connectivity such as the inverse/connectivity measures, the number of scalp electrodes, the head model, the toolboxes used to perform the analysis and intra/inter subject reproducibility of the identified networks [14, 38, 56]. Indeed, the network measures and other issues such as the preprocessing techniques for instance should necessary be also the subjects of further validation/investigation. More precisely, the field needs studies that thoroughly evaluate graph theoretic approaches in combination with different inverse solution and connectivity measures.

## VII. OUTLOOK

As long as there will be technological progress in the EEG systems on one hand and progress in signal processing on the other hand, there will be always new information to extract from the EEG.

In this article, we present one of the latest advances in identifying brain networks, with high spatiotemporal resolution, from dense-EEG recordings: EEG source connectivity. We provide an overview of this approach and present main processing aspects of a signal problem consisting of estimating brain networks at the level of neuronal sources from surface EEG recordings.

We also review the applications of this new 'neuroimaging' technique, in the context of normal brain functions and brain disorders. It is worth mentioning that this review is not exhaustive. The emphasis is on the fundamental aspects of a new neuroimaging technique that provides a good time/space resolution to identify functional brain networks. A number of issues have not been dealt as our intent is to provide a didactic guide for researchers interested in EEG source connectivity. By pointing out some methodological issues, our intent is also to help these researchers to choose/design the methods able to extract relevant information from EEG data in a given application context.

What's next? The signal processing community is directly involved in the new advances mainly in the development of fully automatic pre-processing algorithms, more realistic inverse solutions algorithms and unbiased effective connectivity measures. Efforts will likely lead to the development of novel signal processing methods able to assess the dynamics of the brain networks (on short and long time scales). At the same time, the rapid progress in the network analysis community will certainly improve existing methods for analyzing the brain networks identified from the dense-EEG.

The recent trends in open source neuroimaging data will undoubtedly accelerate the validation of the methods such as the huge database of human connectome project (HCP) http://www.humanconnectome.org/ . The MEG HCP data could be used to test new methods and validate existing methods. In addition, the structural connectome from HCP (mainly the diffusion tensor imaging- DTI- data) could be certainly used as a constraint in the inverse solutions which could lead to an improvement of the spatial precision of the identified functional networks.

ACKNOWLEDGMENT

This work has received a French government support granted to the CominLabs excellence laboratory and managed by the National Research Agency in the "Investing for the Future" program under reference ANR-10-LABX-07-01. It was also financed by the Rennes University Hospital (COREC Project named conneXion, 2012-14 and BrainGraph, 2015-17). Authors would like to thank O. Dufor for helping in the recording of real data used in figure 4 and I. Merlet for helping in the generation of simulations used in figure 5 and 6.